\documentstyle[aps]{revtex}

\begin{document}

\title{Two-dimensional weak localization effects in high temperature 
superconductor Nd$_{2-x}$Ce$_x$CuO$_{4-\delta}$}

\author{G.~I.~Harus, A.~N.~Ignatenkov, A.~I.~Ponomarev, L~.D.~Sabirzyanova, 
N~.G.~Shelushinina}
\address{Institute of Metal Physics\\ 18 Kovalevskaya St., Ekaterinburg, 620219,
Russia}

\author{A.~A.~Ivanov}
\address{Moscow Engineering Physics Institute, Moscow, 115409, Russia}

\maketitle

\begin{abstract}
A systematic study of the resistivity and Hall effect in single crystal 
Nd$_{2-x}$Ce$_x$CuO$_{4-\delta}$ films ($0.12\le x\le 0.20$) is presented, 
with special emphasis on the low-temperature dependence of the normal state 
conductance. Two-dimensional weak localization effects are found both in 
a normally conducting underdoped sample ($x = 0.12$) and {\it in situ} 
superconducting optimally doped ($x = 0.15$) or overdoped ($x = 0.18$) samples 
in a high magnetic field $B > B_{c2}$. The phase coherence time 
$\tau_{\varphi} (5.4 \times 10^{-11}$\,s\ at 2\,K) and the effective thickness 
of a CuO$_2$ conducting layer $d (\cong 1.5$\AA) have been estimated by 
fitting 2D weak localization theory expressions to magnetoresistivity data 
for magnetic fields perpendicular to the $ab$ plane and in plane. Estimates 
of the parameter $d$ ensure strong carrier confinement and justify a model 
consisting of almost decoupled  2D  metallic  sheets for the  
Nd$_{2-x}$Ce$_x$CuO$_{4-\delta}$ single crystal.

\vspace{0.2in}


\end{abstract}

\section{Introduction}
The field of high-transition-temperature (high-$T_c$) superconductivity has 
generated several thousand publications in the last few years. For a short 
overview of the lattice structure and phase diagram of the most widely studied 
copper oxide compounds, such as hole-doped La$_{2-x}$Sr$_x$CuO$_4$ and 
YBa$_2$Cu$_3$O$_{6+x}$ or electron-doped L$_{2-x}$Ce$_x$CuO$_4$ (L=Nd or Pr), 
one can consult, e.g., the review in Ref. [1] or book by Plakida [2]. 
The copper oxide high-$T_c$ materials are basically tetragonal, and all of 
them have one or more CuO$_2$ planes in their structure, which are separated 
by layers of other atoms (Ba-O, La-O, Nd-O, ... ). Most researchers empirically 
and theoretically have come to a consensus that superconductivity is related 
to processes occurring solely in the conducting CuO$_2$ planes, with the other 
layers simply providing the carriers (they are therefore called charge 
reservoirs). In the CuO$_2$ planes, each copper ion is strongly bonded to four 
oxygen ions separated by approximately 1.9\,\AA.

Due to the layered character of the crystal structures, the high-$T_c$ copper 
oxide compounds are highly anisotropic in their normal-state electrical 
properties. Although the resistivity in the CuO$_2$ planes, $\rho_{ab}$, shows 
metallic temperature dependence, the temperature behavior and the magnitude of 
the resistivity parallel to the $c$ axis, $\rho_c$, are strongly dependent on 
crystal structure, and on the concentration of charge carriers.

Systematic data for $\rho_c$ in a number of high-$T_c$ materials obtained on 
well characterized single crystals are presented by Ito et al. [3]. For 
hole-doped systems YBa$_2$Cu$_3$O$_{6+x}$ and La$_{2-x}$Sr$_x$CuO$_4$ $\rho_c$ 
exhibits a marked change in magnitude as well as in temperature dependence 
with changing hole concentration (i.e., changing $x$). For the underdoped 
(low $x$) and optimally doped (superconducting) compounds $\rho_c$ is 
non-metallic ($d\rho_c/dT < 0$) at low enough temperatures. In both systems 
the anisotropy coefficient, $\rho_c/\rho_{ab}$, decreases noticeably with 
doping, being $\sim 10^2$ for the superconducting compounds.

The crystal structure T' of the electron doped Nd$_{2-x}$Ce$_x$CuO$_{4-\delta}$ 
system is the simplest among the superconducting cuprates with the perovskite 
structure [4]. The Cu atoms in the CuO$_2$ layers of hole-doped 
La$_{2-x}$Sr$_x$CuO$_4$ or Nd$_{2-x-y}$Ce$_x$Sr$_y$CuO$_{4-\delta}$ ($y > x$) 
superconductors are surrounded by apical O atoms, forming octahedrons 
(T structure) or semioctahedrons (T$^*$ structure). The most important 
difference in the crystal structures of Nd$_2$(Ce)CuO$_4$ and 
La$_2$(Sr)CuO$_4$ is that the apical oxygen atoms in the T' structure are 
displaced so as to make an isolated CuO$_2$ plane (Fig.1).

The undoped system Nd$_2$CuO$_4$ is an insulator, with the valence band mainly 
of O $2p$ character, and the empty conduction band being the upper Hubbard Cu 
$3d$ band. The Coulomb $3d-3d$ repulsion at the Cu site $U (\cong 6\,-\,7$ eV) 
is strong, and it is greater than the oxygen to metal charge-transfer energy 
$\Delta (\cong 1\,-\,2$ eV). As the gap between the conduction and valence 
bands is determined just by the energy $\Delta$, these cuprates are classified 
as charge-transfer semiconductors [5].

The combination of Ce doping and O reduction results in $n$-type conduction in 
CuO$_2$ layers in Nd$_{2-x}$Ce$_x$CuO$_{4-\delta}$ single crystals [4,6]. 
An energy band structure calculation [7] shows that the Fermi level is located 
in a band of $pd\sigma$-type formed by $3d\,(x^2-y^2)$ orbitals of Cu and 
$p_{\sigma}(x,y)$ orbitals of oxygen. The $pd\sigma$ band appears to be of 
highly two-dimensional (2D) character, with almost no dispersion in the 
$z$-direction normal to CuO$_2$  planes. The electrons are concentrated within 
the confines of conducting CuO$_2$ layers, separated from each other by a 
distance $c\cong 6$\AA.

In accordance with such a structure Nd$_{2-x}$Ce$_x$CuO$_{4-\delta}$, single 
crystals have a significantly higher resistive anisotropy than Y- or 
La- systems: $\rho_c/\rho_{ab}\cong 10^4$ for $x = 0.15$ [8,9] and for\\ 
$x=0.16\,-\,0.20$\ with different values of oxygen deficiency $\delta$ [10] 
at room temperature, it increases with decreasing temperature [10]. 
Measurements by Ito et al. [3] for another electron-doped compound with the 
same T' structure, Pr$_{2-x}$Ce$_x$CuO$_{4-\delta}$, show that for $x~=~0.15$\ 
the anisotropy is very large, $\rho_c/\rho_{ab}\ge 10^4$, and non-metallic 
$\rho_c$\ is observed. Preliminary measurements on a Pr-system with different 
$x$\ indicated that, as in the case of Y- and La- systems, $\rho_c$\ decreases 
with increasing carrier concentration much more rapidly than $\rho_{ab}$.

The larger anisotropy in Nd- or Pr- systems compared with La- or Y- systems 
would imply that fluorite-type Nd$_2$O$_2$ or Pr$_2$O$_2$ layers block 
out-of-plane conduction more effectively than NaCl- type La$_2$O$_2$ or 
Ba$_2$O$_2$ layers [3].

The non-metalic behavior of out-of-plane conductance suggests that the 
carriers are confined tightly in the CuO$_2$ plane [3]. It is thus of interest 
to search for two-dimensional effects in the in-plane conductance of the 
layered cuprates. There are several previous reports on the manifestation of 
2D weak localization effects in the in-plane conductance of 
Nd$_{2-x}$Ce$_x$CuO$_{4-\delta}$ single crystals or films. Thus a linear 
dependence of resistivity on $ln{T}$ comes about at $T < T_c$ for samples 
with $x\cong 0.15$, in which the superconducting state is destroyed by a 
magnetic field [11]. Furthermore, a highly anisotropic negative 
magnetoresistance, predicted for 2D weak localization, has been observed in 
the nonsuperconducting state at low temperatures: in samples with $x = 0.11$ 
[12] and in unreduced samples with $x = 0.15$ [13] or $x = 0.18$ [14]. We 
report here a systematic study of 2D weak localization effects for a number 
of optimally reduced samples of Nd$_{2-x}$Ce$_x$CuO$_{4-\delta}$ with 
$0.12\le x\le 0.20$.

\section{Experimental procedure}
The flux separation technique was used for Nd$_{2-x}$Ce$_x$CuO$_{4-\delta}$ 
film deposition [15]. High-quality $c$-axis oriented single crystal films with 
thickness around 5000\,\AA\ and $0.12\le x\le 0.20$ were grown. The values of 
$T_c$ after sample reduction are shown in Table 1.

Figure~2 demonstrates that $T_c$ of the film with $x = 0.15$ is in agreement 
with previously published data for bulk single crystals [4]. The values of 
$T_c$ for overdoped films with $x\ge 0.17$ are higher than for corresponding 
bulk crystals in accordance with the information of Xu {\it et al.} [16] that 
superconductivity survives up to $x = 0.22$ in Nd$_{2-x}$Ce$_x$CuO$_{4-\delta}$ 
films.

Standard four-terminal measurements of the resistivity $\rho$ and Hall effect 
$R\ (\vec{j}||ab, \vec{B}||c)$ in the dc regime were carried out in the 
temperature range $T_c < T\le 300$\,K without a magnetic field $B$, and in 
magnetic fields up to $B = 12$\,T at temperature down to 0.2\,K. The electrical 
contacts were prepared by evaporating thin silver strips onto the sample, and 
attaching silver wires to these with conducting glue.

\section{Results}
The temperature dependence of the zero-field in-plane resistivity for the 
investigated samples at T up to 300\,K is shown in Fig.\,3. A clear resistance 
minimum is observed at $T\cong 150$\,K for the nonsuperconducting sample 
with $x = 0.12$. The $\rho(T)$ dependence is described by 
$\rho = \rho_0 + A\cdot T^2$ at $T = (20 - 180)$\,K for $x$ = 0.15, 0.17, and 
0.18, and over the wider $T = (10 - 300)$\,K for $x$ = 0.20. The values of 
$\rho_0$ and $A$ are presented in Table 1.

We describe here the magnetoresistance measurements only for the underdoped 
nonsuperconducting sample ($x$ = 0.12) and for two overdoped superconducting 
samples ($x$\,=\,0.18 and $x$ = 0.20). Detailed investigations of $\rho(B,T)$ 
dependencies for the optimally doped sample with $x$ = 0.15 were presented 
earlier [17], as were some results on the $x$ = 0.18 sample [18].

In the superconducting samples, normal-state transport at low $T$ is hidden 
unless the magnetic field stronger than the second critical field $B_{c2}$ 
is applied. As we are interested in the low-temperature $\rho(T)$ 
dependence, we have destroyed the superconductivity with a magnetic field 
$B_{\perp}$ perpendicular to the CuO$_2$ planes. In  Fig.~4, the 
$\rho(B_{\perp})$ dependence for $x$\,=\,0.20 at $T$\,=\,1.3\,K and $T$\,=\,4.2\,K 
in a magnetic fields up to $B$ = 5.5\,T are presented. In the inset of Fig.~4, 
the dependence of the Hall coefficient $R$ on magnetic field $B_{\perp}$ at 
$T$\,=\,1.3\,K is also shown. On the assumption that $B_{c2}^{\perp}(T)$ is 
a field in which $\rho(B_{\perp})$ and $R(B_{\perp})$ at given $T$ come up to 
their normal-state value, we have $B_{c2}^{\perp}$\,=\,2.2\,T at $T$\,=\,1.3\,K 
and $B_{c2}^{\perp}$\,=\,1.5\,T at $T$\,=\,4.2\,K.

In our previous investigation [18] of the sample with $x$ = 0.18, negative 
magnetoresistance was observed after the destruction of superconductivity 
by a magnetic field up to 5.5\,T at $T\ge$ 1.4\,K. In Fig.~5 $\rho(B_{\perp})$ 
is shown for this sample at much lower temperatures (down to 0.2\,K) and in 
fields up to 12\,T. The inset of Fig.~5 shows an example of $R(B_{\perp})$ 
at given $T < T_c$. The nonmonotonic $R(B_{\perp})$ behavior, with reversal 
of the sign of the Hall effect, is usually observed in the mixed state of the 
superconductor [18,19]. The transition to the normal state is completed at 
$B = B_{c2}$, where the Hall coefficient becomes nearly constant with the 
same value as in the normal state at $T > T_c$ ($B_{c2}\cong 5$\,T at 
$T$ = 0.2\,K). Values of $B_{c2}$ for $x$ = 0.18 estimated in this way at 
different temperatures are marked by the arrows in Fig.~6. This figure also 
clearly demonstrates the transition from positive to negative 
magnetoresistance after the suppression of superconductivity.

In Fig.~7, the results of the theoretical description of the 
magnetoconductivity at $B > B_{c2}$ are presented. Figure~8 demonstrates 
that the resistivity of the sample with $x$ = 0.18 is a linear function of 
$\ln T$\ in magnetic fields $B > B_{c2}$. The experimental points for 
$B$ = 1.5\,T are also shown. If the logarithmic temperature dependence of 
the resistivity is typical of the normal state, then the discrepancy between 
the experimental points and the logarithmic law indicates that the normal 
state has not yet been attained at $B$ = 1.5\,T.

We have also measured the in-plane conductivity in nonsuperconducting sample 
with $x$\,=\,0.12 for magnetic fields perpendicular and parallel to the CuO$_2$ 
planes up to 5.5\,T at $T$\,=\,1.9\,K and 4.2\,K (Fig.~9). The positive 
magnetoconductivity (negative magnetoresistance) observed in this sample is 
obviously anisotropic relative to the direction of the magnetic field.

\section{Discussion}
A logarithmic temperature dependence of the conductivity is one indication 
of the interference quantum correction due to 2D weak localization. A magnetic 
field normal to the diffusion path of a carrier destroys the interference. 
In a two-dimensional system, it causes negative magnetoresistance for the 
field perpendicular to the plane, but no effect for the parallel 
configuration. Weak localization effects are almost totally suppressed for 
$B_{\perp} > B_{tr}$ [20], where the so called ``transport field'' is defined 
as the field at which 
\begin{equation}
2\pi B_{tr}\ell^2 = \Phi_0.
\end{equation}
Here $\ell$ is the elastic mean free path and $\Phi_0 = \pi c\hbar/e$ is the 
elementary flux quantum.

Let us compare Eq. (1) with the relations between the coherence length 
$\xi$ and the second critical field in the pure superconductor ($\xi\ll\ell$),
\begin{equation}
2\pi B_{c2}\xi^2 = \Phi_0,
\end{equation}
or in the so-called ``dirty limit'' ($\xi\gg\ell$):
\begin{equation}
2\pi B_{c2}\xi\ell = \Phi_0.
\end{equation}
From Eqs. (1) and (2) we have $B_{tr}/B_{c2} = (\xi/\ell)^2$, so 
$B_{tr}\ll B_{c2}$, and it is impossible to observe weak localization effects 
in the pure case. In contrast, from Eqs. (1) and (3) one has 
$B_{tr}/B_{c2} = (\xi/\ell)$, $B_{tr} > B_{c2}$, and weak localization should 
survive in the normal state ($B > B_{c2}$) of a dirty superconductor.

In Table 2, the parameters of investigated samples essential to a description 
of localization are presented. From the experimental values of the in-plane 
resistivity $\rho$ and Hall constant $R$\ in the normal state, we have obtained 
the surface resistance $R_s=\rho/c\ $\ per CuO$_2$\ layer and the bulk and 
surface electron densities $n=(eR)^{-1}$\ and\ $n_s=n\!\times\!c$\ \ 
($c$ = 6\,\AA\ is the distance between CuO$_2$ layers). Using the relations 
[21] $\sigma_s = (e^2/\hbar)k_F\ell$\ for the 2D conductance 
$\sigma_s = 1/R_s$, and $k_F = (2\pi n_s)^{1/2}$\ for the Fermi wave vector, 
we have estimated the important parameter $k_F\ell$, the mean free path $\ell$, 
and then according to Eq. (1) the characteristic field $B_{tr}$. For the 
sample with $x$ = 0.15  we use the data of Ref. [17].

In a random two-dimensional system, the parameter $k_F\ell$ can serve as a 
measure of disorder [21]. It is seen from Table 2 that we have a wide range 
$\sim (1 - 10^2)$\ of $k_F\ell$\ in the investigated series of samples. For 
$k_F\ell\gg 1$, a true metallic conduction takes place in CuO$_2$ layer. 
Thus, we have a rather pure 2D system with $k_F\ell\sim 10$\ for $x$ = 0.15 
or $x$ = 0.18, and an extremely pure system with $k_F\ell\sim 10^2$\ for 
$x$ = 0.20. It is quite remarkable that even at such high values of the 
parameter $k_F\ell$, a trace of localization comes to light: for 
$B_{\perp} > B_{c2}$, $\rho$\ is greater at 1.3\,K than at 4.2\,K (see Fig.~4).
As for the sample with $x$ = 0.12, where $k_F\ell$\ is of the order of unity, 
this system is in close proximity to transition from weak logarithmic to strong 
exponential localization as disorder increases ($k_F\ell$\ decreases).

The second critical field $B_{c2}^{\perp}$\ at temperatures around 
$T \cong$ 1.4\,K (see Figs. 4 and 5) and values of $\xi$\ estimated according 
to Eqs. (2) or (3) are also shown in Table 2. In the pure system with 
$x$\,=\,0.20, $\xi < \ell$, and negative magnetoresistance is not detected at 
$B > B_{c2}$, at least for $T \ge$ 1.3\,K (see Fig. 4). Systems with 
$x$ = 0.15 and 0.18 are situated close to the dirty limit $\xi\gg\ell$, and 
there exist appreciable field ranges $B_{c2} < B < B_{tr}$\ where negative 
magnetoresistance due to 2D weak localization is actually observed (see 
Ref. [17] and Fig.~6).

In 2D weak localization theory, the quantum correction to the Drude 
conductivity in a perpendicular magnetic field is [22]:
\begin{equation}
\Delta\sigma_s(B_{\perp})=\alpha\frac{e^2}{2\pi^2\hbar}
\biggl\{\Psi\Bigl(\frac{1}{2}+\frac{B_{\varphi}}{B_{\perp}}\Bigr)-
\Psi\Bigl(\frac{1}{2}+\frac{B_{tr}}{B_{\perp}}\Bigr)\biggr\}  
\end{equation}
where $\alpha$\ is a factor of the order of unity, $\Psi$\ is the digamma 
function, and $B_{\varphi} = c\hbar/4eL_{\varphi}^2$. Here 
$L_{\varphi}=\sqrt{D\tau_{\varphi}}$\ is the phase coherence length, $D$ is 
the diffusion coefficient, and $\tau_{\varphi}$ is the phase breaking time.

The fit of (4) to the experimental $\sigma_s(B_{\perp})$ data for $x$ = 0.18 
at $B_{\perp} > B^{\perp}_{c2}$ is shown in Fig.~7. For each temperature there 
are two fit parameters: the characteristic field $B_{\varphi}$ (or 
$L_{\varphi}$) and the factor $\alpha$. The widest range of requisite magnetic 
fields and thus the most accurate fit results are obtained for the lowest 
temperature $T$ = 0.2\,K. With $B_{tr}$ = 22.5\,T, the best fit is obtained 
for $B_{\varphi}\cong$ 0.1\,T ($L_{\varphi}$ = 560\,\AA) and $\alpha$ = 6.6. 
The fitting procedure is highly sensitive to the value of the parameter 
$\alpha$. In contrast, the value of $B_{\varphi}$\ is obtained only to order 
of magnitude, as we have no zero-field and weak-field data. Nevertheless, 
there is no doubt that the inequality $B_{\varphi}\ll B_{tr}$ is valid.

In the field range $B_{\varphi}\ll B\ll B_{tr}$, the expression (4) can be 
written
\begin{equation}
\Delta\sigma_s(B_{\perp})=\alpha\frac{e^2}{2\pi^2\hbar}
\biggl\{-\Psi\Bigl(\frac{1}{2}\Bigr)-
\ln \frac{B_{\perp}}{B_{tr}} \biggr\}.
\end{equation}
The inset of Fig.~7 shows that the experimental data at $T$ = 0.2\,K can be 
fitted rather closely by this simple formula over a wide range of fields, 
5\,T $\le B\le$\ 11\,T. But as we have the factor $\alpha$ = 6.6, negative 
magnetoresistance is too large to be due to the destruction of weak 
localization only. Thus we conclude that some additional mechanism of negative 
magnetoresistance must be at work.

There exists another quantum correction to the normal-state conductivity with 
a logarithmic dependence of magnetoresistivity on $B$, namely the correction 
due to disorder-modified electron-electron interaction (EEI) in the Cooper 
channel [23]. In the range of magnetic fields $B_T\ll B\ll B_{tr}$, we have
\begin{equation}
\Delta\sigma_s^{EEI}(B_{\perp})=-\frac{e^2}{2\pi^2\hbar}
\cdot g(T)\cdot \ln \biggl(\frac{B_{\perp}}{B_{tr}} \biggr),
\end{equation}
where $B_T = \pi c\hbar/2eL_T^2,\ L_T = \sqrt{\hbar D/kT}$ is the thermal 
coherence length, and $g(T)$ is the effective interaction constant of two 
electrons with opposite momenta. For the attractive electron-electron 
interaction due to virtual phonon exchange, $g > 0$, and according to (6) the 
magnetoresistance should be negative.

As we have dealt with {\it in situ} superconducting samples, so that $g > 0$, 
the contribution due to EEI is most probably the reason for the extra negative 
magnetoresistance at very low temperatures ($B_T$ = 0.02\,T at $T$ = 0.2\,K). 
With increasing temperature, the magnitude of the EEI conribution decreases 
rapidly ($\alpha$ = 2.5 at $T$ = 0.8\,K), and at $T \ge$ 1\,K the estimated 
value of the factor is close to unity ($\alpha$ = 0.77 at $T$ = 2.1\,K), as 
it should be for weak localization.

It should be noted that pronounced negative magnetoresistance due to the 
suppression of weak electron localization is observed in ordinary 
superconductors as well. The electron transport properties of three-dimensional 
(3D) amorphous $\alpha$-Mo$_3$Si and $\alpha$-Nb$_3$Ge superconducting films 
have been investigated in magnetic fields up to $B$ = 30\,T at temperatures 
down to $T$ = 0.35\,K [24]. The authors have found that both the temperature 
and field dependence of the resistivity $\rho$ can be qualitatively described 
by weak localization theory. At low temperatures and in magnetic fields above 
the upper critical field, $B > B_{c2}$, magnetoconductivity is proportional 
to $B^{1/2}$. This field dependence is consistent with weak localization in 
the high-field limit ($B \gg B_{\varphi}$) for 3D disordered systems, in 
contrast to a 2D system with $\Delta\sigma(B)\sim \ln B$.

One important indication of the 2D character of a system is the strong 
dependence of magnetoresistance on magnetic field orientation. Highly 
anisotropic (negative) magnetoresistance is actually observed in a 
nonsuperconducting sample with $x$ = 0.12 (see Fig. 9). From the fit to 
$\sigma_s(B_{\perp})$\ by the functional form (4) with $\alpha$\,=\,1, we find 
$L_{\varphi}$\,=\,770\,\AA\ at $T$\,=\,1.9\,K and $L_{\varphi}$\,=\,550\,\AA\ 
at $T$\,=\,4.2\,K, so that the phase coherence time 
$\tau_{\varphi}\,=\,5.4\times 10^{-11}$\,s\ at $T$\,=\,1.9\,K and 
$\tau_{\varphi}\,=\,2.7\times 10^{-11}$\,s\ at $T$\,=\,4.2\,K.

We explain the much weaker negative magnetoresistance for the parallel 
configuration $B\| ab$ by incorporating finite-thickness ($d$) corrections 
into the strictly 2D theory [25]:
\begin{equation}
\Delta\sigma_s(B_{\|})=\frac{e^2}{2\pi^2\hbar}
\cdot \ln \Biggl(1+\frac{d^2L_{\varphi}^2}{3\lambda_{\|}^4} \Biggr),
\hspace{0.5in} \lambda_{\|}^2=\frac{c\hbar}{eB_{\|}}.
\end{equation}
By fitting the theoretical expression (7) to the curves for $\sigma(B_{\|})$ 
(see Fig. 9), we have found the effective thickness of a conducting CuO$_2$ 
layer, $d\cong$ 1.5\,\AA. This value yields an estimate for the extent 
of the electron wave function in the normal direction, and ensures strong 
carrier confinement ($d < c$). The single crystal NdCeCuO can therefore be 
regarded as an analog of an ultra-short-period superlattice 
(1.5\,\AA wells / 4.5\,\AA barriers).

As the 2D version of weak localization theory is able to describe the behavior 
of $\sigma(B,T)$\ in our sample, the inequality $\tau_{esc} > \tau_{\varphi}$ 
should be valid for the escape time of an electron from one CuO$_2$ plane to 
another. Then we have  $\tau_{esc} \ge 5\times 10^{-11}$\,s. The escape time 
between adjacent wells in a superlattice can also be estimated from the value 
of the normal diffusion constant, $\tau_{esc} = c^2/D_{\perp}$. For the 
parameters of our sample at 300\,K, we have [9] 
$D_{\|}/D_{\perp} = 1.7\times 10^4$\ with the in-plane diffusion constant 
$D_{\|}$ = 1.2 cm$^2$\,s$^{-1}$. Then $\tau_{esc} \cong 5\times 10^{-11}$\,s 
even at room temperature, so $\tau_{esc} > \tau_{\varphi}$\ with certainty at 
low temperatures. 

\section{Conclusion}
We have investigated the low-temperature and magnetic field dependence of the 
normal state in-plane resistivity, $\rho_{ab}$, in a layered copper oxide 
single crystal Nd$_{2-x}$Ce$_x$CuO$_{4-\delta}$. The material is regarded as 
an intrinsic two-dimensional conduction system (a collection of 2D conducting 
CuO$_2$\ planes), and the results are interpreted in terms of the 2D weak 
localization model. Three indications of 2D weak localization have been 
displayed: logarithmic temperature dependence of  the resistivity, 
significant negative magnetoresistance for a field normal to the $ab$-plane, 
and pronounced magnetoresistance anisotropy (much weaker effect for a parallel 
configuration). A strong dependence of the magnitude of magnetoresistance on 
the direction of the magnetic field is the most important experimental test 
for the two-dimensional character of a conducting system.

In a series of samples with $x$\,=\,(0.12\,-\,0.20), we have a full range of 
disorder parameter values, $k_F\ell$\,=\,(2\,-\,150). Estimates of essential 
microscopic parameters, such as the elastic mean free path $\ell$, the 
inelastic scattering length $L_{\varphi}$, and the effective thickness of a 
conducting layer $d$, have shown that in accordance with the adopted model, 
$d\ll \ell$\ < $L_{\varphi}\ll t$\ ($t$ is the geometrical thickness of a 
sample). Moreover, our estimates show that the thickness of the conducting 
layer is less than the distance between CuO$_2$\ layers, $d\,<\,c$, and this 
favors carrier confinement within a separate CuO$_2$\ layer. Thus, the NdCeCuO 
single crystal can be described as a natural superlattice with a confining 
potential induced both by the specific $pd\sigma$\ symmetry of the electron 
wave function and strong Coulomb correlation effects.

\vspace{0.2in}

This research was supported by the Russian Program ``Current Problems in 
Condensed Matter Physics'' (Grant  No.98004) and the Russian Foundation for 
Basic Research (Grant No.99-02-17343).

\newpage
Table 1.
\\
\vspace{0.2in}

\begin{tabular}{|c|c|c|c|c|c|c|}                                                                                                 \hline
$x$  & $t$, \AA & $T_c$, K & $\rho_0\times 10^5$, & $\rho_{300\,K}\times 10^5$, & $\rho_{300\,K}/\rho_0$ & $A\times 10^9$,  \\   
     &          &          & ohm$\cdot$cm         & ohm$\cdot$cm                &                        & ohm$\cdot$cm/K$^2$ \\ \hline
0.12 & 5500     & -        & -                    & 102                         & -                      & -   \\                \hline
0.15 & 5000     & 20       & 8.2                  & 42.4                        & 5.2                    & 4.0 \\                \hline
0.17 & 5700     & 12       & 8.6                  & 29.6                        & 3.4                    & 2.7 \\                \hline
0.18 & 5000     & 6.0      & 6.0                  & 23.5                        & 3.9                    & 2.2 \\                \hline
0.20 & 4000     & $<\,1.3$ & 1.1                  & 10.0                        & 9.1                    & 1.1 \\                \hline
\end{tabular}
\\
\vspace{0.5in}

Table 2.
\\
\vspace{0.2in}

\begin{tabular}{|c|c|c|c|c|c|c|} \hline
$x$  & $n\times 10^{-22}$, cm$^{-3}$ & $k_F\ell$ & $\ell$, \AA & $B_{tr}$, T & $B_{c2}^{\perp}$, T & $\xi$, \AA \\ 
     &                               &           &             &             & ($T$ = 1.4\,K)      &            \\ \hline
0.12 & 0.2$^*$ & 2   & $\sim 10$  & $\sim 270$  & -   &  -  \\ \hline
0.15 & 2.0$^*$ & 18  & $\sim 30$  & $\sim 30$   & 5.5 & 180 \\ \hline
0.18 & 1.1     & 25  & 40         & 22.5        & 4.0 & 200 \\ \hline
0.20 & 1.0     & 150 & 240        & 0.6         & 2.2 & 150 \\ \hline
\end{tabular}
\\

$^*$ Data from Ref. [4] at $T$ = 80 K.

\vspace{0.2in}
\begin{center}
Figures
\end{center}

Fig. 1. Crystal structure of three types of copper oxides (Ref. [4]).

Fig. 2. Phase diagram of Nd$_{2-x}$Ce$_x$CuO$_{4-\delta}$. Notation: 
triangles and circles - data of Ref. [4] 
(triangles - $T_c$ = 0); crosses - our data.

Fig. 3. Temperature dependence of in-plane resistivity for the samples 
investigated Nd$_{2-x}$Ce$_x$CuO$_{4-\delta}$.

Fig. 4. In-plane resistivity ($j \perp B$) of the sample with $x$ = 0.20 as 
a function of magnetic field $B \perp ab$\ at two different temperatures. 
Arrows indicate values of the second critical field. Inset : Hall coefficient 
($j\| ab; B\perp ab$) as a function of magnetic field at $T$ = 1.3\,K.

Fig. 5. In-plane resistivity ($j \perp B$) of the sample with $x$ = 0.18 as 
a function of magnetic field $B \perp ab$\ at different temperatures. 
Inset: Hall coefficient ($j\| ab; B \perp ab$) as a function of magnetic field 
at $T$ = 0.2\,K. The arrow shows the estimate for the second critical field.

Fig. 6. Negative magnetoresistance at $B > B_{c2}$\ in the sample with 
$x$ = 0.18. Arrows indicate values of the second critical field $B_{c2}$\ at 
different temperatures. 

Fig. 7. Fit of the expression (4) to experimental data on the surface 
conductivity of the sample with $x$\,=\,0.18 at $T$\,=\,0.2\,K. Fit parameters 
of the broken line are: $B_{\varphi}$\,=\,0.1\,T, $\alpha$\,=\,6.6. Inset: 
Surface conductivity as a function of $\ln B$.

Fig. 8. In-plane resistivity of the sample with $x$ = 0.18 as a function of 
$\ln T$\ in different magnetic fields.

Fig. 9. Surface conductivity of the sample with $x$ = 0.12 as a function of 
magnetic field  $B_{\perp}$\ ($B \perp ab$) or $B_{\|}\ (B \| ab)$\ at 
different temperatures.


\begin{thebibliography}{29}

\protect\bibitem{_1} E.~Dagotto, {\it Rev.~Mod.~Phys.} {\bf 66}, 763 (1994).

\protect\bibitem{_2} N.~M.~Plakida "High-temperature superconductors", Moscow, 
International Program of Education (1996).

\protect\bibitem{_3} T.~Ito, Y.~Nakamura, H.~Takagi, and S.~Uchida, 
{\it Physica C} {\bf 185-189}, 1267 (1991);\\
T.~Ito, H.~Takagi, S.~Ishibashi T.~Ido and S.~Uchida, 
{\it Nature} {\bf 350}, 596 (1991).

\protect\bibitem{_4} S.~Uchida, H.~Takaji and Y.~Tokura, 
{\it ISEC-89}, Tokyo, 306 (1989).

\protect\bibitem{_5} J.~Zaanen, G.~A.~Sawatsky and J.~W.~Allen, 
{\it Phys. Rev. Lett.} {\bf 55}, 418 (1985).

\protect\bibitem{_6} H.~Takagi, S.~Uchida and Y.~Tokura, 
{\it Phys. Rev. Lett.} {\bf 62}, 1197 (1989). 

\protect\bibitem{_7} S.~Massidda, N.~Hamada, J.~Yu. and A.~J.~Freeman,
{\it Physica C} {\bf 157} 571 (1989).

\protect\bibitem{_8} Z.~Z.~Wang, T.~R.~Chien, N.~R.~Ong et al., 
{\it Phys. Rev. B} {\bf 43} 3020 (1991).

\protect\bibitem{_9} A.~I.~Ponomarev, V.~I.~Tsidilkovski, K.~R.~Krylov et al., 
{\it Journal of Superconductivity} {\bf 9}, 27 (1996).

\protect\bibitem{_10} Beom-hoan~O and J.~T.~Markert, 
{\it Phys. Rev. B} {\bf 47}, 8373 (1993).

\protect\bibitem{_11} Y.~Hidaka, Y.~Tajima, K.~Sugiama et al. 
{\it J.Phys. Soc. Jap.} {\bf 60}, 1185 (1991).

\protect\bibitem{_12} S.~J.~Hagen, X.~Q.~Xu, W.~Jiang et al., 
{\it Phys. Rev. B} {\bf 45}, 515 (1992).

\protect\bibitem{_13} A.~Kussmaul, J.~S.~Moodera, P.~M.~Tedrow et al., 
{\it Physica C} {\bf 177}, 415 (1991).

\protect\bibitem{_14} S.~Tanda, M.~Honma and T.~Nakayama, 
{\it Phys. Rev. B} {\bf 43}, 8725 (1991).

\protect\bibitem{_15} A.~A.~Ivanov, S.~G.~Galkin, A.~V.~Kuznetsov et al., 
{\it Physica C} {\bf 180}, 69 (1991).

\protect\bibitem{_16} X.~Q.~Xu, S.~N.~Mao, Wu~Jiang et al., 
{\it Phys. Rev. B} {\bf 53}, 871 (1996).

\protect\bibitem{_17} A.~I.~Ponomarev, K.~R.~Krylov, G.~I.~Harus et al.,
{\it JETP Lett.} {\bf 62}, (1995).
 
\protect\bibitem{_18} G.~I.~Harus, A.~N.~Ignatenkov, N.~K.~Lerinman et al.,
{\it JETP Lett.} {\bf 64}, 444 (1996).

\protect\bibitem{_19} Y.~X.~Jia, J.~Z.~Liu, M.~D.~Lan et al., 
{\it Phys. Rev. B} {\bf 47}, 6043 (1993).

\protect\bibitem{_20} M.~I.~Dyakonov, 
{\it Sol. St. Commun.} {\bf 92}, 711 (1994).

\protect\bibitem{_21} P.~A.~Lee and T.~V.~Ramakrishnan,
{\it Rev. Mod. Phys.} {\bf 57}, 287 (1985).

\protect\bibitem{_22} S.~Hikami, A.~Larkin and Y.~Nagaoka, 
{\it Progr. Theor. Phys.} {\bf 63}, 707 (1980).

\protect\bibitem{_23} B.~L.~Altshuler, A.~G.~Aronov, A.~P.~Larkin, 
D.~E.~Khmelnitskii, {\it JETP} {\bf 81}, (1981).

\protect\bibitem{_24} A.~V.~Samoilov, N.-C.~Yeh and C.~C.~Tsuei, 
{\it Phys. Rev. B} {\bf 57}, 1206 (1998).

\protect\bibitem{_25} B.~L.~Altshuler and A.~G.~Aronov, 
{\it JETP Lett.} {\bf 33}, (1981).

\end{thebibliography}
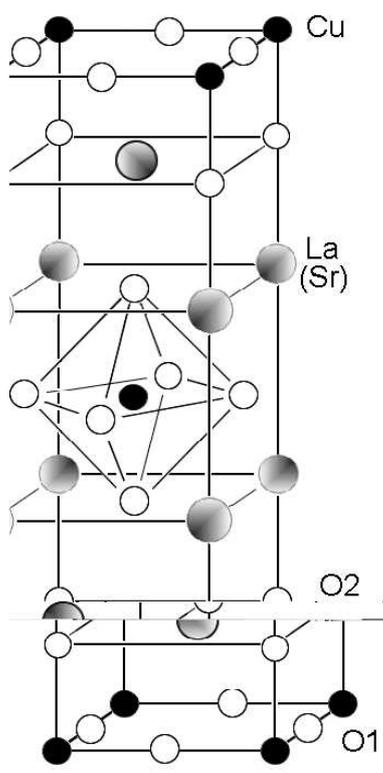
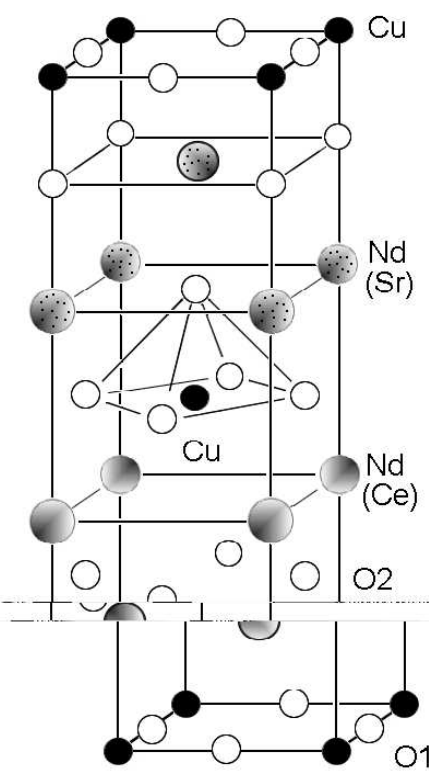
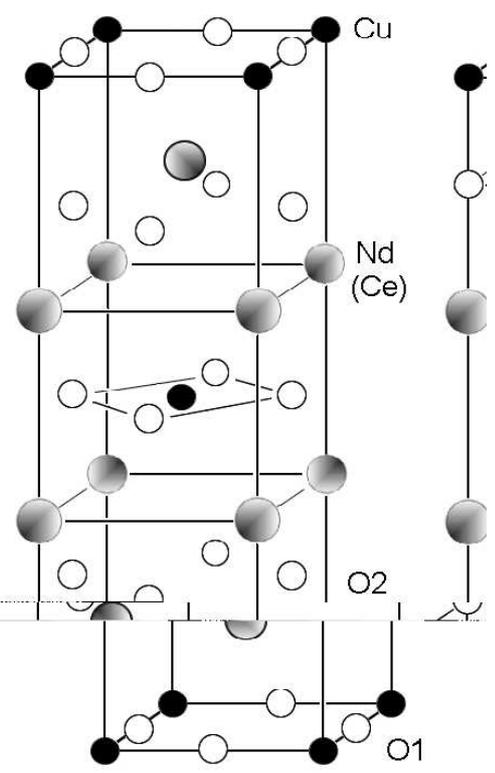
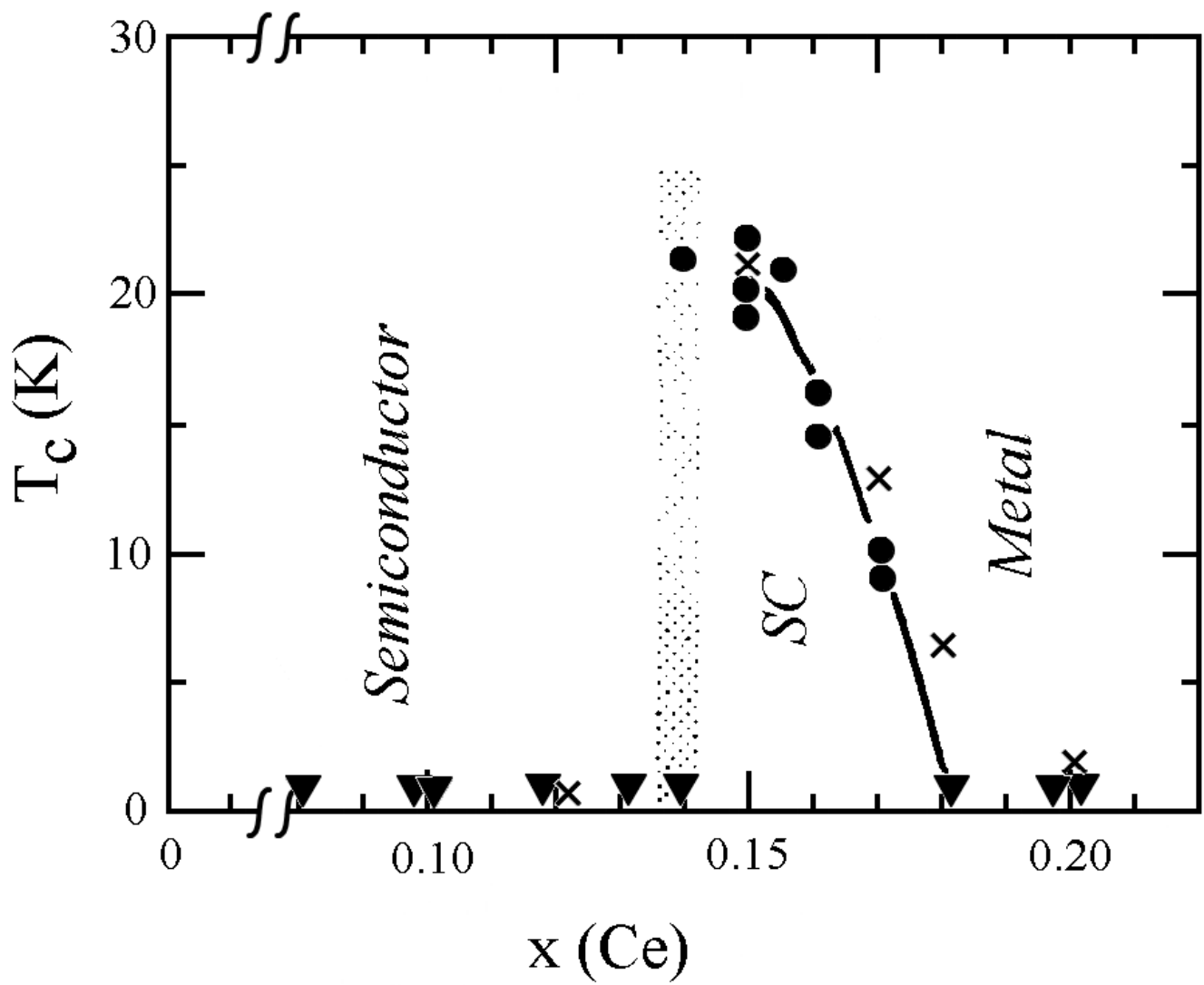
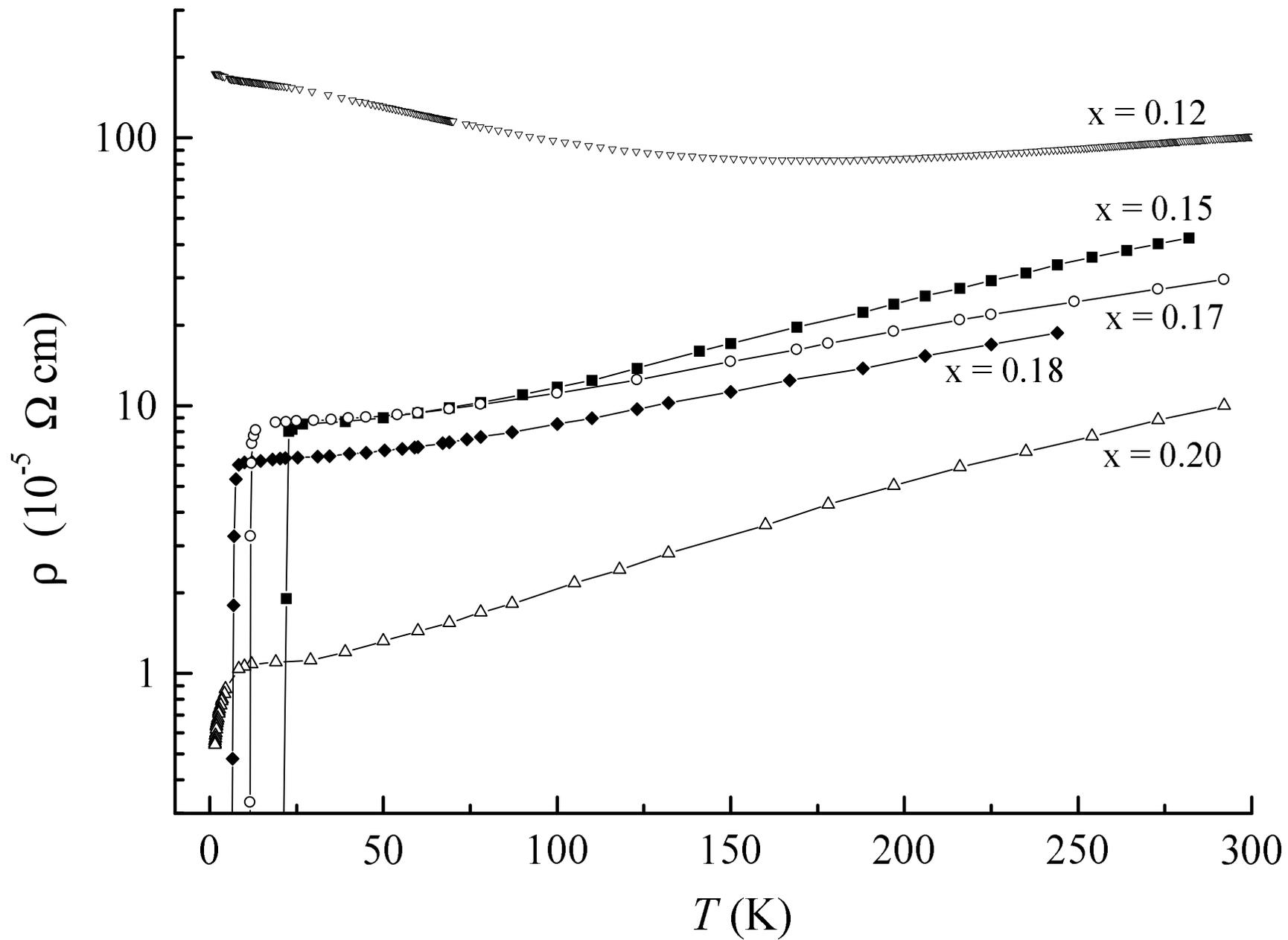
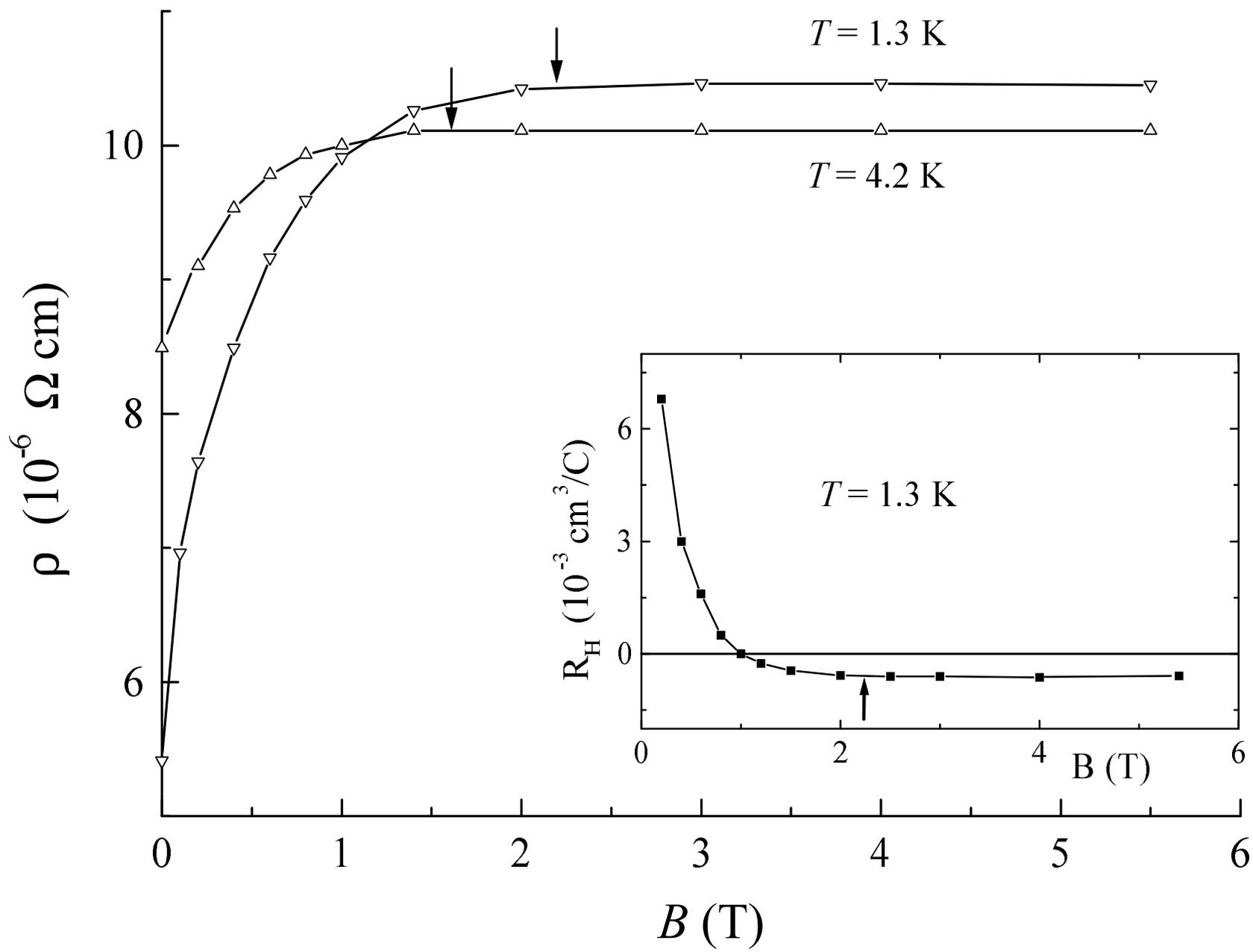
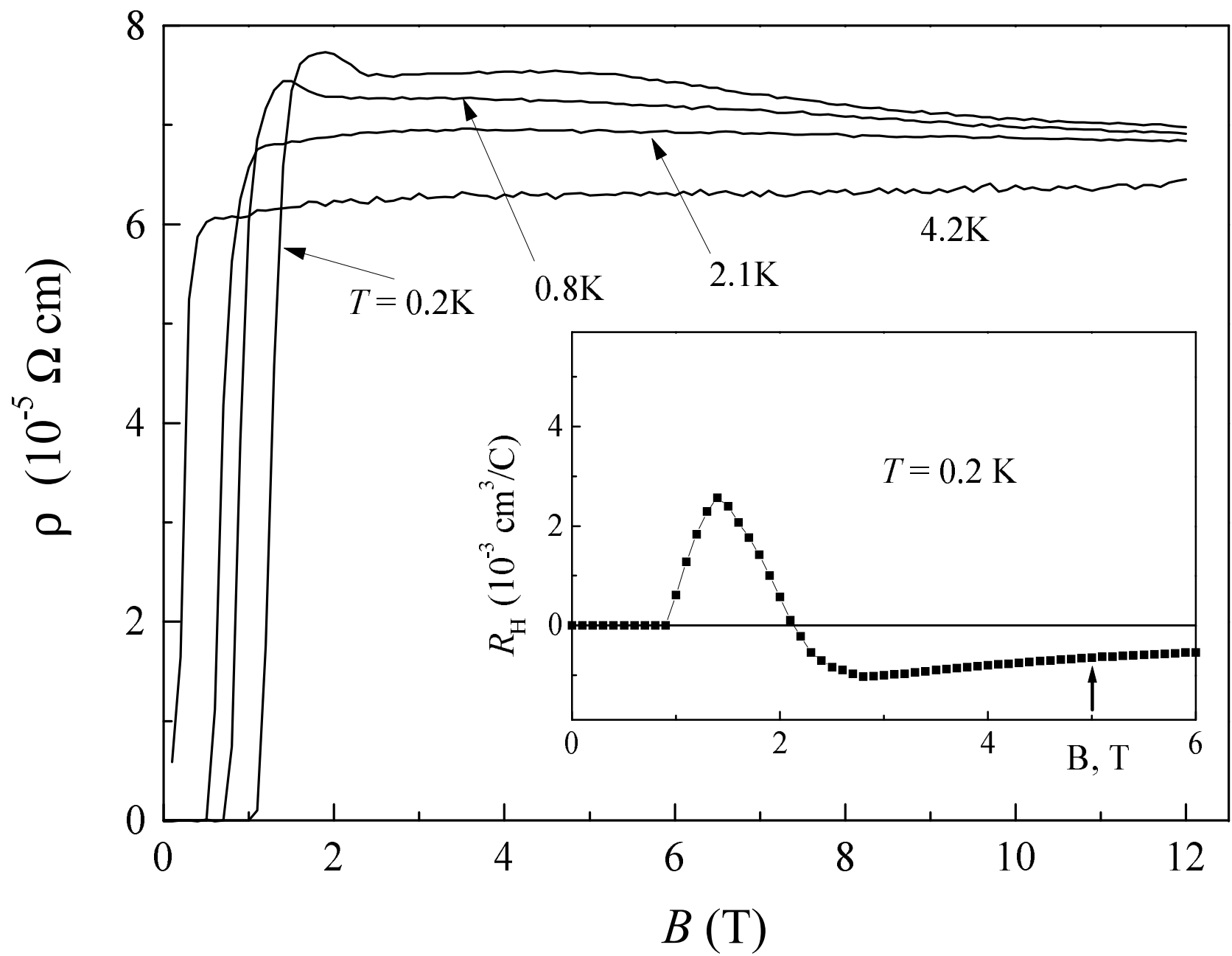
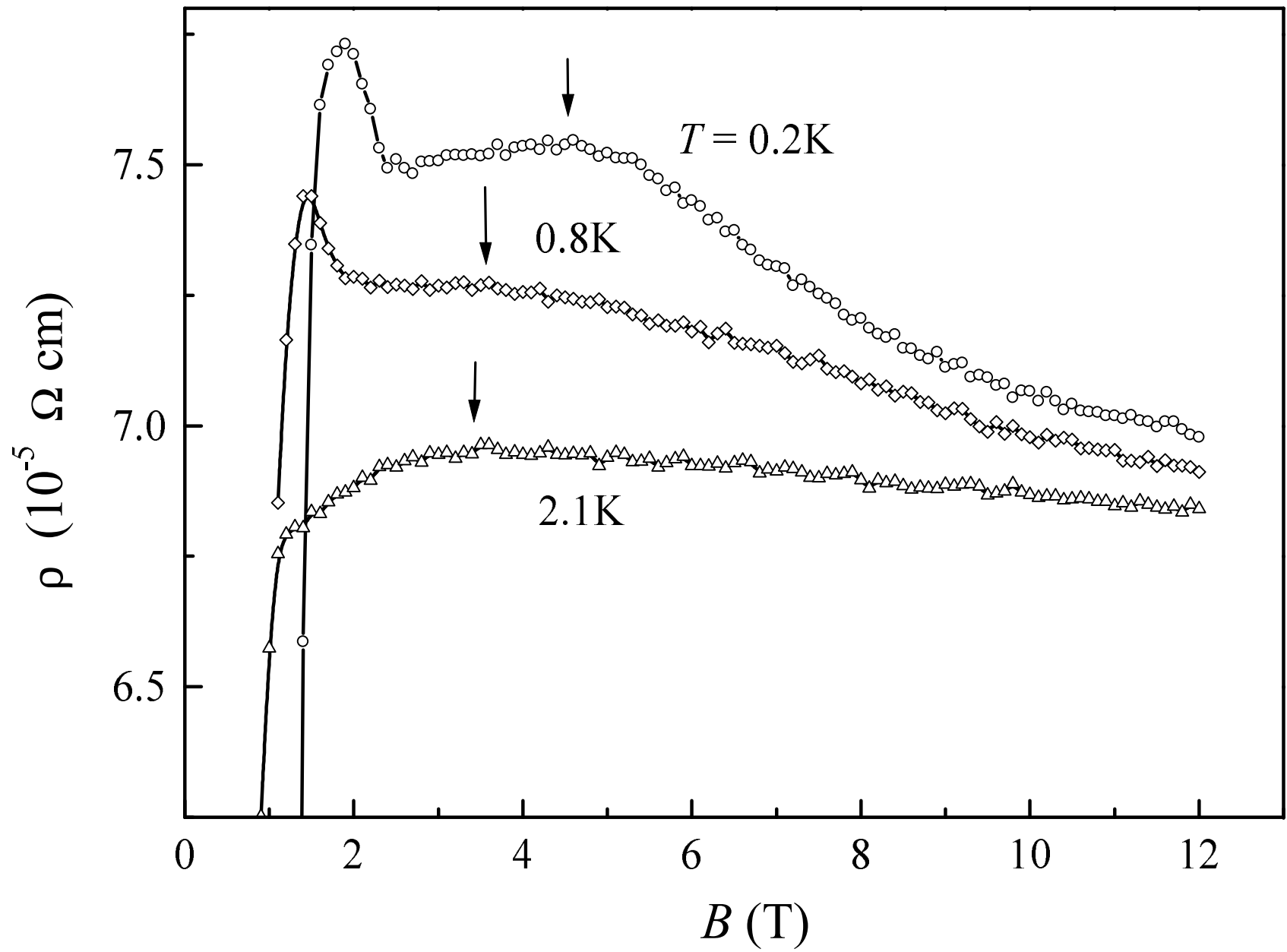
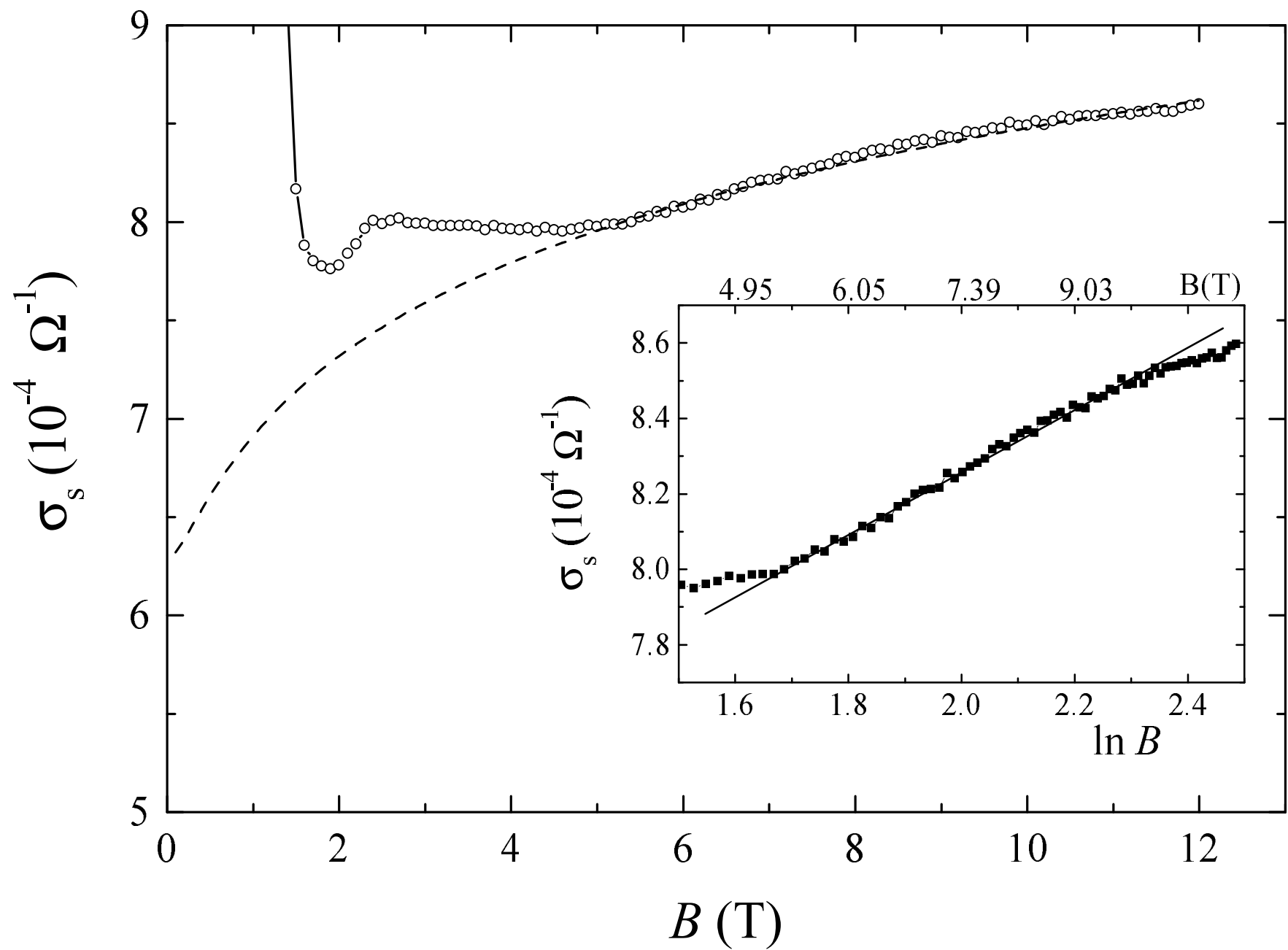
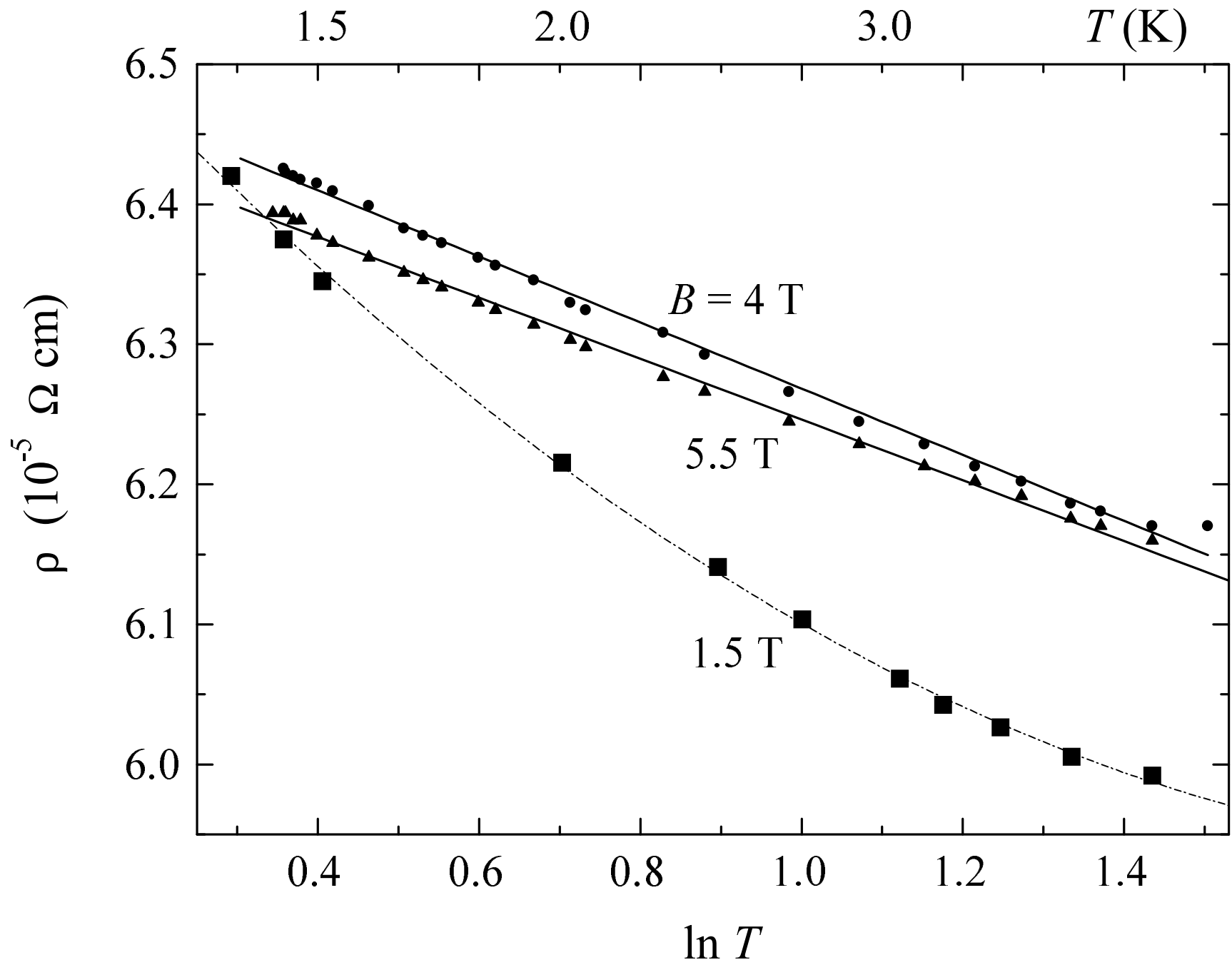
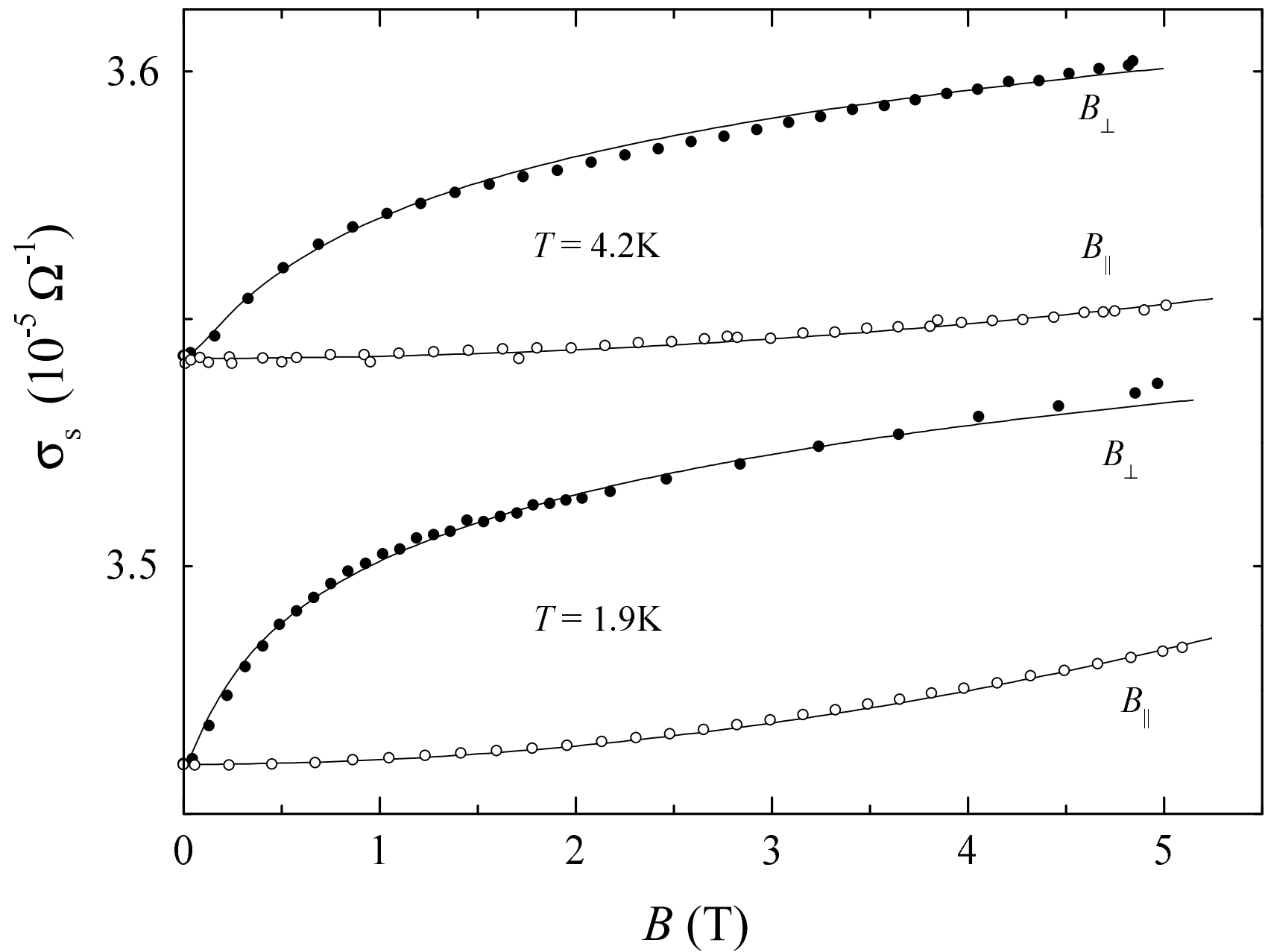
\end{document}